\documentclass{llncs}
\input{preambule.sty}
\usepackage{subfigure} 
\usepackage{placeins}
\usepackage{graphicx,wrapfig,lipsum} 
\usepackage[rightcaption]{sidecap}
\usepackage[titletoc,title]{appendix}
\title{Diffeomorphic Multi-Resolution Deep Learning Registration for Applications in Breast MRI}

\author{Matthew G. French\inst{1} \and
Gonzalo D. Maso Talou\inst{1} \and
Thiranja P. Babarenda Gamage\inst{1} \and
Martyn P. Nash\inst{1,2} \and 
Poul M. Nielsen\inst{1, 2} \and 
Anthony J. Doyle\inst{3} \and
Juan Eugenio Iglesias\inst{4, 5} \and
Ya\"el Balbastre\inst{4} \and
Sean I. Young\inst{4, 5}}

\institute{
Auckland Bioengineering Institute, University of Auckland, NZ.
\and
Department of Engineering Science and Biomedical Engineering, University of Auckland, NZ. 
\and  
Faculty of Medical and Health Sciences, University of Auckland, NZ.
\and
MGH/HST Martinos Center for Biomedical Imaging, Harvard Medical School,
Boston, MA, USA. 
\and Computer Science and Artificial Intelligence LAB (CSAIL), MIT, Cambridge, MA, USA.}

\begin{document}

\maketitle

\begin{abstract} 
In breast surgical planning, accurate registration of MR images across patient positions has the potential to improve the localisation of tumours during breast cancer treatment.
While learning-based registration methods have recently become the state-of-the-art approach for most medical image registration
tasks, these methods have yet to make inroads into breast image registration due to certain difficulties—the lack of rich texture information in breast MR images and the need for the deformations to be diffeomophic. In this work, we propose learning strategies for breast MR image registration that are amenable to diffeomorphic constraints, together with early experimental results from  in-silico and in-vivo experiments. One key contribution of this work is a registration network which produces superior registration outcomes for breast images in addition to providing diffeomorphic guarantees. 
% We hope that this work and the accompanying dataset will spur general interest in the breast MR image registration problem and encourage the MICCAI community to engage in this less explored area of medical imaging.
\end{abstract}

\section{Introduction} 
\begin{figure}[t]
    \centering
    \includegraphics[width=\textwidth]{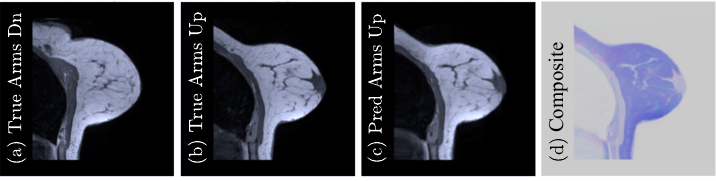}
    \caption{Breast MR Image Registration (viewed in a transverse plane). We propose a diffeomorphic registration-based approach to localise in-vivo breast tissue between multiple positions. An arms down image (a) is registered to an arms up image (b) in the prone position. We show our predicted arms up image (c) along with the false colour composite of the predicted (magneta), ground-truth (cyan) images, depicting purple in the regions of agreement (most of the tissue) (d).}
    \label{fig:teaser}
\end{figure}

Globally, breast cancer is the most diagnosed cancer for women, contributing to 11.7\% cancer incidence rates \cite{Sung2021}. Dynamic Contrast-Enhanced Magnetic Resonance Imaging (DCE-MRI) is commonly used for detecting breast cancer in women with a high risk of developing breast cancer. This imaging is performed in the prone position to reduce breathing artifacts. Breast Conservation Therapy (BCT) is the most common treatment for patients with early-stage breast cancer. BCT involves localisation of the cancer lesion in the supine position, followed by lumpectomy (excision of the lesion) and often radiotherapy to eliminate any residual disease. The success of BCT depends on the accurate localisation of tumours inside the breast in the supine position. The deformation between prone and supine can vary significantly depending on e.g. breast size, tissue density, age, etc. Carbonaro et al.  \cite{Carbonaro2012} reports a median lesion displacement between prone and supine breast MRI as ranging from 30-60mm. Such lesions can range from very small ($<$10mm) to very large ($>$60mm), however 20mm is the most common size at which breast cancer is diagnosed \cite{Spoik2018,Narod2013}. Due to breasts intrinsic high-lipidic composition, the non-linear stress-strain relationship of the skin (which highly restricts tissue deformation) and changes in arm positioning, the breast tissue exhibits a large 
 and complex deformation between the diagnostic and pre-operative positions. This makes tracking tumours between the prone and supine positions extremely challenging. Computed Tomography imaging is typically used to guide radiotherapy, however, such intra-operative image guidance is typically not available during lumpectomy. 

Techniques for localising tumour positions during lumpectomy, e.g. using guide wires, fail in outlining the tumour in its entirety. Such challenges may contribute to 20–40\% reoperation rates reported in the literature \cite{Pleijhuis2009}. Previous studies proposed the acquisition of an additional pre-operative supine MRI to overcome the challenges of tumour localisation \cite{Alderliesten2010}. However, due to respiratory artifacts influencing the image quality, clinicians do not generally acquire contrast-enhanced images in the supine position. Respiratory artifacts in the breast tissue are absent in the prone position as the patients chest is fixed relative to the coils, i.e. the coils are positioned around the breast and the torso against the MR table. In the supine position, even positioning coils against the breast (e.g. air blanket coils) will produce respiratory artifacts as the posterior region of the torso is constrained by the MR table and the expansion of the torso will mainly be manifested in the anterior region. Therefore tumour locations still need to be identified from diagnostic DCE-MRI in the prone position, and  mapped to the proposed pre-operative non-contrast MRI in the supine position. This is challenging due to the large and complex deformations that the breast undergoes between these positions, limiting the clinical applicability of this approach. Developing robust techniques to map the breast tissue between the diagnostic and pre-operative positions can potentially improve BCT outcomes by providing accurate tumour localisation. 

In this work, we propose a learning-based, diffeomorphic registration method for localising breast tissue across  positions (Fig. 1). At the core of our registration method is SVFlowNet, a novel Stationary Velocity Field (SVF)-based registration approach that is constrained to retrieve diffeomorphic transformations. Our results show a better performance in comparison  to state-of-the-art non-diffeomorphic deep learning approaches on two breast MRI datasets.
\begin{enumerate}
    \item We extend the dual-stream network architecture \cite{kang2022} to diffeomorphic registration to provide diffeomorphic guarantees.
    \item We introduce a differentiable SVF composition layer based on the BCHD (or  Baker–Campbell–Hausdorff–Dynkin) formula \cite{Mger2019NotesOT}.
    \item We evaluate the effect of supervision strategy and diffeomorphic encoding on the accuracy of breast MR image registration.
\end{enumerate}

% Our main contributions are: 

% \begin{enumerate}
%     \item 
% \end{enumerate}

\section{Related Work} 
In recent years, there has been an effort to map breast deformations between diagnostic and pre-operative positions using a combination of computational biomechanics and medical image registration techniques \cite{Han2014,Hipwell2016,BabarendaGamage2017,Mira2018,BabarendaGamage2019}.

While biomechanics approaches show promise, limitations exist in their ability to recover the large deformations the breast can undergo. For example, accounting for the change in relative positions of the pectoral muscles and the base of the breast (deep, superficial fascia) as the individual and their arms change position between the diagnostic and pre-operative positions is a challenge that has not yet been addressed with biomechanical modelling. This results in the shoulder joint and the arms rolling posteriorly, stretching the pectoral muscles, and flexing the ribcage, resulting in complex breast deformation. Developing methods to quantify and understand these complex deformations would help identify approaches to improve predictions from biomechanical models and enable their application for navigational guidance when 3D imaging is unavailable e.g. during surgical interventions.  

Image registration is a deeply nonlinear and nonconvex problem, which has been historically solved using iterative methods. The seminal work of Lucas and Kanade \cite{lucas1981} and Horn and Schunk \cite{HornSchunck1981} showed that a linearization 
of the equations of motion leads to a linear relationship between the temporal gradients of two images and the motion flow. 
While this linearization forms the basis of all gradient-descent registration algorithms, it only holds true in the small deformation regime. More advanced deformation models have been proposed to solve large-deformation registration, such as the large deformation diffeomorphic metric mapping (LDDMM) framework \cite{beg2005} and its log-Euclidean 
variant \cite{Arsigny2006}, which assumes SVFs. Both approaches ensure that the resulting deformations are diffeomorphic, and therefore one-to-one and onto.

% This linearization forms the basis of all gradient-descent registration algorithms, which then 
% differ mostly in terms of similarity metric, regularizer, and optimization strategy. A final difference lies in 
% the encoding of the spatial transform itself. It was rapidly found that even with regularization, dense displacement 
% fields estimated by gradient descent can lose their invertibility. This is especially true when the target displacement is very large or 
% convoluted. A solution was found in the field of differential equations: the transformation can be broken down into  a series of small transformations, whose composition can be large while still being invertible. In the limit of infinitely many steps, these small transformations are the temporal velocities of the deformation flow. This led to the large deformation diffeomorphic metric mapping (LDDMM) framework \cite{beg2005}, and its simplified variant---the log-Euclidean 
% framework---that assumes that all small deformations are equal (i.e., the velocities are ``stationary'') \cite{Arsigny2006}. Invertibility of the deformation are highly desirable in breast MR image registration prevents self-folding (tissue loss).

Iterative approaches have now been superseded by learning-based approaches \cite{Balakrishnan2019}. Inspired by iterative registration, a number of works have investigated pyramidal representations of the flow field \cite{Mok2020,Li2022,kang2022,Young2022}. However, \cite{kang2022,Young2022} do not enforce bijectivity, and while \cite{Mok2020,Li2022} encode deformations using SVFs, they do not take advantage of the properties of the Lie algebra when combining flows across scales. In this work, we propose a principled extension of the dual-stream architecture from \cite{kang2022,Young2022} that properly handles SVFs.

\section{Computational Framework}  
Accurately aligning breast MR images relies heavily not only on the proposed registration network  (SVFlowNet) but also on the learning strategies (supervised and unsupervised) and loss functions used. We discuss each of these in turn.

\subsection{Constructing SVFlowNet}  

\textbf{Flow U-Net Architecture.}  SVFlowNet extends   Flow U-Net   \cite{Young2022}, which forms the basis of our approach and serves as our non-diffeomorphic baseline. Flow U-Net \cite{Young2022} proposes two major modifications to the U-Net architecture to form a dual stream pyramid network for registration (see Appendix A). The work of \cite{Young2022} propagates the deformations via addition $\phi^{(l + 1)} = \psi_\textrm{up}(\phi^{(l - 1)}) + \phi^{(l)}$, which is a first-order approximation of the composition. We extend the work of Young et al. \cite{Young2022}, by propagating the deformations via  composition (linear interpolation) to avoid unnecessary error introduced by deformation addition. We denote the convolutions that extract the flow and perform upsampling by $\psi_\textrm{conv}$ and $\psi_\textrm{up}$, respectively. We implement the upsampling ($\psi_\textrm{up}$) operator as linear interpolation between the resolution at layers $l$ and $l+1$.
\\

% \begin{enumerate}
%     \item \textbf{Flow blocks:} consisting of a sequence of convolution blocks which are tasked with extracting the deformations $\phi^{(l)}$ at each U-Net decoder layer $l$ from the corresponding features maps $(f_0^{(l)}, f_1^{(l)}\circ \psi_\textrm{up}(\phi^{(l - 1)}))$ where $\psi_\textrm{up}$ upsamples the deformations to the next resolution using linear interpolation. The flow blocks also upsample the previously extracted deformation $\phi^{(l - 1)}$, which is propagated through the decoder by computing the composition mapping $\phi^{(l + 1)} = \psi_\textrm{up}(\phi^{(l - 1)}) \circ \phi^{(l)}$. The work of \cite{Young2022} propagates the deformations via addition $\phi^{(l + 1)} = \psi_\textrm{up}(\phi^{(l - 1)}) + \phi^{(l)}$ which is a first-order approximation of the composition. Here we have extended their work by propagating the deformations via the composition mapping (linear interpolation) to avoid unnecessary error introduced by deformation addition. We denote the convolutions that extract the flow and perform upsampling by $\psi_\textrm{conv}$ and $\psi_\textrm{up}$, respectively. We implement the upsampling ($\psi_\textrm{up}$) operator as linear interpolation between the resolution at layers $l$ and $l+1$. 

% \end{enumerate} 

\noindent \textbf{Baker–Campbell–Hausdorff–Dynkin Layers.} SVFlowNet parameterises the multi-resolution output of the Flow U-Net flow blocks \cite{Young2022} as SVFs. The SVF at each resolution $\mathbf{v}^{(l)}$ are integrated via scaling and squaring to obtain the corresponding deformation $\phi^{(l)}$ which by construction is diffeomorphic. With the addition of the SVF parameterisation of the flow block output and the {scaling and squaring} (denoted by $\exp$), a SVF at the layer $l$ can be expressed as follows
\begin{align}
    \mathbf{v}^{(l-1)'} & = \psi_\textrm{up} \left(\mathbf{v}^{(l-1)}\right)\\
    \phi^{(l-1)} & = \exp\left(\mathbf{v}^{(l-1)'}\right)\\
    \mathbf{v}^{(l)'} & = \psi_\textrm{conv}^{(l)} \left( H \left(f_0^{(l)}, f_1^{(l)} \circ \phi^{(l-1)}\right) \right) \\
    \mathbf{v}^{(l)} & = \zeta\left(\mathbf{v}^{(l-1)}, \mathbf{v}^{(l)'}\right) ~,
\end{align}
in which the $\zeta$ operator denotes the series expansion resulting from the work of Baker, Campbell, Hausdorff and Dynkin (BCHD) \cite{Mger2019NotesOT}.
% \begin{align}\label{eqn:BCHD}  
%     \nonumber
%     &\zeta: (\mathbf{v}^{(l-1)}, \mathbf{v}^{(l)}) \mapsto \mathbf{v}^{(l-1)} + \mathbf{v}^{(l)} + \frac{1}{2}[\mathbf{v}^{(l-1)}, \mathbf{v}^{(l)}] ... \\ ... + \frac{1}{12}([\mathbf{v}^{(l-1)}, & [\mathbf{v}^{(l-1)}, \mathbf{v}^{(l)}]] + 
% [\mathbf{v}^{(l)}, [\mathbf{v}^{(l)}, \mathbf{v}^{(l-1)}]]) - \frac{1}{24}[\mathbf{v}^{(l)}, [\mathbf{v}^{(l-1)}, [\mathbf{v}^{(l-1)}, \mathbf{v}^{(l)}]]] ...
% \end{align}  
% where 
% \begin{align}
%     [\mathbf{v}^{(l-1)}, \mathbf{v}^{(l)}] &= \mathbf{v}^{(l-1)} \circ \mathbf{v}^{(l)} - \mathbf{v}^{(l)} \circ \mathbf{v}^{(l-1)}
% \end{align}
% are the Lie brackets under the composition mapping $\circ$. 
In the series limit, the BCHD operator ensures
\begin{align}
    \exp(\zeta (\mathbf{v}^{(l-1)}, \mathbf{v}^{(l)})) &= \exp(\mathbf{v}^{(l-1)}) \circ \exp(\mathbf{v}^{(l)});
\end{align} 
see \cite{Mger2019NotesOT} for details.  

The operator $\zeta$ enables the implicit propagation of deformations through the explicit propagation of SVFs, avoiding integration error introduced by {scaling and squaring} $\mathbf{v}$ to obtain $\phi = \exp(\mathbf{v})$. Additionally, $\zeta$ propagation accommodates for the propagation of both non-commutative and commutative multi-resolution SVFs. This can be verified by observing that by construction the BCHD formula yields propagation via summation ($\zeta: (\mathbf{v}^{(l - 1)}, \mathbf{v}^{(l)}) \mapsto  \mathbf{v}^{(l-1)} + \mathbf{v}^{(l)} $) for the case where $(\mathbf{v}^{(l - 1)}, \mathbf{v}^{(l)})$ commute. This is a theoretical improvement on the work of \cite{Mok2020} and \cite{Li2022}, who propose propagation via summation. In this work will implement $\zeta$ as the BCHD series truncated after the fourth order term (see Appendix B).
\\

\noindent \textbf{Hadamard Transform.} As in Flow U-Net \cite{Young2022}, we reparametrise the flow block feature input using $H: (f_0^{(l)}, f_1^{(l)'}) \mapsto (f_0^{(l)} + f_1^{(l)'}, f_0^{(l)} - f_1^{(l)'})$, which can be thought of as the Hadamard transform \cite{Pratt1969} of features across channels. It should be noted that we have let $f_1^{(1)'} = f_1^{(l)} \circ \psi_\textrm{up}(\phi^{(l - 1)})$ in this case only, to avoid complicating the expression which has been introduced. 
% It is hypothesized that the Hadamard transform reparameterisation helps the flow blocks as $(f_0^{(l)}, f_0^{(l)} - f_1^{(l)'})$ must be processed to extract the deformation via Horn \& Schuncks equation \cite{HornSchunck1981} and $f_0 + f_1$ of the Hadamard transform is motivated by median filtering techniques in optical flow \cite{Sun2014}.

\subsection{Learning Strategies}  
In this work, we consider both supervised and unsupervised learning approaches to optimize Flow U-Net and the SVFlowNet variants over the in-silico dataset. For the in-vivo task we use the unsupervised learning approach as the ground-truth deformation is unknown. Consider a deformation field, computed by a neural network $g$ with parameters $\theta$, i.e., $g_{\theta}(f_0, f_1) = \phi$, $\forall (f_0, f_1) \subset \mathcal{D}$ where $\mathcal{D}$ is a given registration dataset. Using this notation, the supervised and unsupervised learning frameworks are defined in the following manner.
\\

\noindent \textbf{Supervised Learning.} The supervised learning approach uses the ground-truth deformations $\hat{\phi} \in \mathcal{D}$ to optimise $\theta$ via 
\begin{equation}\label{eqn:supervised_learning}
    \hat{\theta} = \text{arg} \min_{\theta}\mathcal{L}(g_{\theta}(f_0, f_1), \hat{\phi})
\end{equation}
in which case $\mathcal{L}$ denotes the mean squared error loss  between  predicted $\phi=g_{\theta}$ and ground-truth $\hat{\phi}$ deformations.
\\
% , or equivalently, \begin{equation}\label{eqn:supervised_loss}
%     \mathcal{L}(\phi, \hat{\phi}) = \frac{1}{3|\Omega|} \sum_{\mathbf{x} \in \Omega} || \phi(\mathbf{x}) - \hat{\phi}(\mathbf{x})||_{2}^{2}.
% \end{equation}  

\noindent \textbf{Unsupervised Learning.} The unsupervised learning approach uses the image pair $(f_0, f_1) \in \mathcal{D}$ and the predicted deformation $\phi$ alone to optimise the parameters of the neural network $\theta$ by exploiting the composition mapping $f_1 \circ \phi$ which should approximate $f_0$, i.e. $f_1 \circ \phi \approx f_0$. Minimising the similarity $\mathcal{L}_\textrm{sim}$ between $f_1 \circ \phi$ and $f_0$ alone leads to an ill-posed problem, therefore an additional smoothness term, or regularizer, $\mathcal{L}_\textrm{smooth}$ (weighted by $\lambda \in \mathbb{R}$) is introduced to improve the posedness of the problem. Thus, the training problem in the unsupervised case can be posed as 
\begin{align}\label{eqn:unsupervised_learning}
    \hat{\theta} = \text{arg} \min_{\theta}~ (1 - \lambda)\mathcal{L}_\textrm{sim}(f_1 \circ g_{\theta}(f_0, f_1), f_0) + \lambda \mathcal{L}_\textrm{smooth}(g_{\theta}(f_0, f_1))
\end{align} 
in which $\mathcal{L}_\textrm{sim}(f_1 \circ \phi, f_0)$ denotes the negated normalised cross correlation (NCC) of $f_1 \circ \phi$ and $f_0$; and  
\begin{equation}\label{eqn:unsupervised_loss_smooth}
    \mathcal{L}_\textrm{smooth} = \frac{1}{3 |\Omega|} \sum_{\mathbf{x} \in \Omega} || \nabla^n \mathbf{u}(\mathbf{x})||_{2}^2
    \end{equation}
    is the regularization term. Here, the first-order ($n=1$) gradient is used for the in-silico task and the second-order ($n=2$) gradient is used in the in-vivo task to accommodate for piece-wise linear intensity boundaries.
\\

% \begin{align}\label{eqn:unsupervised_loss_img} 
%     &= -\frac{\big(\sum_{\mathbf{x} \in \Omega} (f_1 \circ \phi(\mathbf{x}) - \mu([f_1 \circ \phi])) (f_0(\mathbf{x}) - \mu(f_0))\big)^2}{\big(\sum_{\mathbf{x} \in \Omega}(f_1 \circ \phi(\mathbf{x}) - \mu([f_1 \circ \phi]))^2 \big) \big( \sum_{\mathbf{x} \in \Omega}(f_0(\mathbf{x}) - \mu(f_0))^2 \big)}
% \end{align}  

\noindent \textbf{Implementation.} Our method is implemented in  PyTorch  \cite{PyTorch} and experiments are performed on an NVIDIA A100 GPU\footnote{https://www.nvidia.com/en-us/data-center/a100/} with 80 GB of memory. Stochastic Gradient Descent with momentum ($\beta = 0.9$) is used as the network optimiser with an initial learning rate of $10^{-2}$. The Reduce On Plateau \cite{PyTorch} learning rate scheduler is applied with a reduction factor of $0.5$. The training is stopped once the learning rate is less than $10^{-6}$. The data is fed to the network in batches of 8 samples. For the in-silico task a ratio of $80/10/10$ is used to split the 1000 samples for training, validation and testing. The test dataset is not used during the learning/optimisation process to determine the optimal parameters of the network.

% \section{Breast MR Image Datasets}   

% \subsection{B-Spline In-silico}  

% \subsection{Arms Up and Down In-Vivo}  

\section{Experimental Results}     
To assess the applicability of SVFlowNet to breast MR image registration, we conduct an extensive quantitative analysis of the deformations produced by SVFlowNet, Flow U-Net \cite{Young2022} and a U-Net \cite{Ronneberger2015,dalca2019unsupervised}, comparing their statistics. We hypothesize that SVFlowNet will achieve the best performance as there is no tissue coming in or out of the field of view in this dataset, and the mechanical deformation will not violate mass conservation, preserving breast tissue between poses. This is guaranteed by a bijective transformation.

% with those from other well-performing registration methods.

\subsection{In-Silico Experiments} 

\begin{figure}[t]
    \centering
    \includegraphics[width=\textwidth]{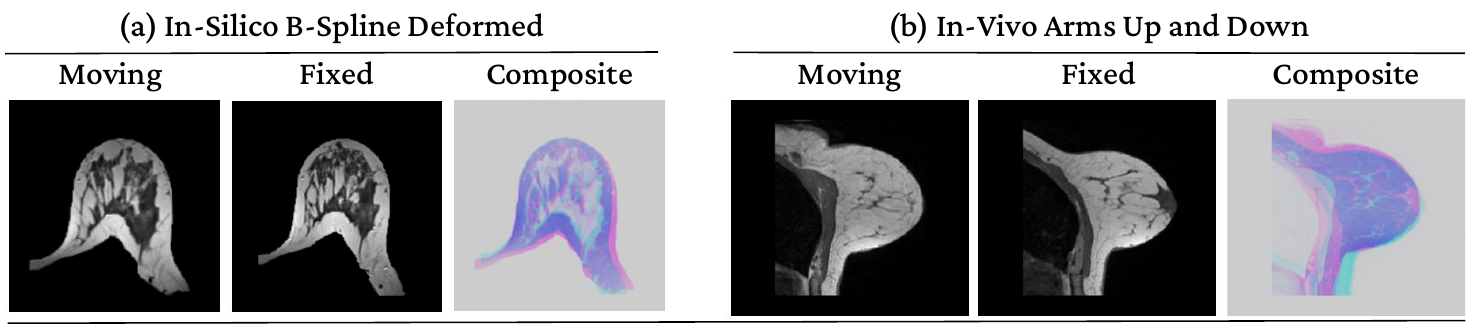}
    \caption{Breast MR Image datasets. (a) and (b) depicted the moving and fixed images from the in-silico and in-vivo datasets, respectively, from the Breast Biomechanics Research Group dataset at Auckland Bioengineering Institute. In each false, colour composite the moving and fixed images are magenta and cyan respectively.}
    \label{fig:breast_datasets}
\end{figure}

Our T2-weighted MRI ($\approx 1 \textrm{mm}^{3}$) in-silico dataset consists of the breast region  of a volunteer in
a prone position which we deform with  $10^{3}$ randomly sampled B-spline based deformations. 
The deformations is generated by B-spline interpolation over the image domain $\mathbf{x} \in \Omega$ via, 
\begin{equation}\label{eqn:bspline_flow}
    \phi_{\text{B-spline}}(\mathbf{x}) = \sum_{\gamma \in \Gamma} \gamma \beta^{(n, D)}(\mathbf{x}).
\end{equation}  
where $\beta^{n}$ is an $n^{th}$ order multi-variate B-spline \cite{Unser1999} over a random grid $\Gamma$ of control points $\gamma$ and $D=3$ is the number of spatial dimensions.  
We use the $5^\textrm{th}$ order B-spline and sample a $3 \times 3 \times 3$ grid of random control points $\gamma \sim \mathcal{N}(0, 1)$. This approach leads to large non-linear deformations that are smooth and approximately meet the incomprehensibility requirements of true breast deformation (see Fig. \ref{fig:breast_datasets}). The mean Jacobian determinant and displacement over the in-silico dataset is $0.95 \pm 0.07$ (local volume change) and $9.56 \pm 3.22 \ (\textrm{mm})$ respectively. 

For the in-silico task we apply unsupervised learning over $\lambda \in (0.1, 0.01, 0.001)$ to characterise the sensitivity of the approach to the regularisation weight. An analysis of the similarity of $\hat{\phi}$ (ground-truth deformation) and optimal deformations $\phi = g_{\hat{\theta}}$ using the SVFlowNet variants and Flow U-Net was performed over the test data using the sum of squared error (SSE) metric to measure flow discrepancy ($\varepsilon_\textrm{flow}$).
% following metrics:  
% \begin{equation}\label{eqn:error_flow}
%     \varepsilon_\textrm{flow}(\hat{\phi}, \phi)  =  \frac{1}{3|\Omega|} \sum_{\mathbf{x} \in \Omega} || \hat{\phi}(\mathbf{x}) - \phi(\mathbf{x})||_{2}^{2};
% \end{equation} 
\\

\noindent \textbf{Supervised.} Considering the median $\varepsilon_\textrm{flow}$, slight improvement can be observed from SVFlowNet for the two SVF propagation techniques (i.e. summation and $\zeta$ propagation) compared to Flow U-Net in the supervised case (Fig. \ref{fig:insilico_eval}). See Appendix C for an ablation study with SVFlowNet variants.   
\\

\noindent \textbf{Unsupervised.} Over all $\lambda$, the $\zeta$-propagation yields the highest accuracy (with $\lambda = 0.1$) on the deformation discrepancy $\varepsilon_\textrm{flow}$, achieving an $\varepsilon_\textrm{flow}$ of $0.021 \pm 0.0074$ voxels (where $\pm 0.0074$ refers to the standard deviation). This is an improvement from Flow U-Net which achieves an $\varepsilon_\textrm{flow}$ of $0.042 \pm 0.0012$ voxels. Furthermore, $\zeta$ propagation out performs summation propagation, as summation achieves an $\varepsilon_\textrm{flow}$ of $0.023 \pm 0.0080$ voxels. However, summation still outperforms both Flow U-Net. This is evidence that SVFlowNet with implicit propagation (summation or $\zeta$) improves the deformation discrepancy results of Flow U-Net. Furthermore, $\zeta$ propagation is shown to be the best-performing propagation technique with respect to the deformation discrepancy (Fig. \ref{fig:insilico_eval}). See Appendix C for an ablation study with variants of SVFlowNet.

% \begin{figure}[htp!]%
% \centering
% \subfigure[Unsupervised]{%
% \label{fig:first}%
% \includegraphics[height=2in]{figures/sensitivity_insilico.png}}%
% \qquad
% \subfigure[Supervised]{%
% \label{fig:second}%
% \includegraphics[height=2in]{figures/super_insilico.png}}%
% \caption{SVFlowNet variants and Flow U-Net performance on the in-silico data assessed using the deformation discrepancy $\epsilon_\textrm{flow}$.} 
% \label{fig:sensitivity_and_super}
% \end{figure}   

% \begin{figure}
%     \centering
%     \includegraphics[width=0.5\textwidth]{figures/insilico_eval.png}
%     \caption{SVFlowNet variants and Flow U-Net performance on the in-silico data assessed using the deformation discrepancy $\epsilon_\textrm{flow}$. }
%     \label{fig:insilico_eval}
% \end{figure}

% \begin{SCfigure}[0.5][ht]
% \caption{SVFlowNet variants and Flow U-Net performance on the in-silico data assessed using the deformation discrepancy $\epsilon_\textrm{flow}$.}
% \includegraphics[width=0.6\textwidth]{figures/insilico_eval.png} 
% \label{fig:insilico_eval}
% \end{SCfigure} 

\begin{figure}[t]
    \centering
    \includegraphics[width=\textwidth]{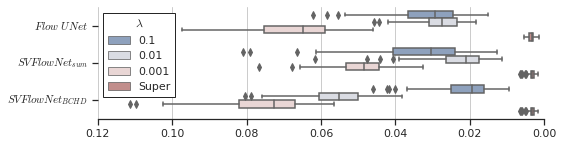}
    \caption{SVFlowNet variants and Flow U-Net performance on the in-silico data assessed using the deformation discrepancy $\epsilon_\textrm{flow}$ (SSE). We show the results of $\epsilon_\textrm{flow}$ (SSE) for supervised learning, and unsupervised learning over a range of regularization weights $\lambda$.}
    \label{fig:insilico_eval}
\end{figure}

% \begin{wrapfigure}{O}{0.25\textwidth}
% \includegraphics[width=6.5cm]{figures/insilico_eval.png} 
% \caption{}\label{fig:insilico_eval}
% \end{wrapfigure} 

\subsection{In-vivo Experiments} 

For the in-vivo task, a pair of breast images, one with arms up and the other with arms down, is manually obtained from a specified region of interest that encompasses all tissues of a single breast (see Figure \ref{fig:breast_datasets}). These images are derived from high-resolution ($1mm^3$) isotropic T1-weighted MR images\footnote{Ethical approval was obtained for this study from the Auckland Health Research Ethics Committee (AH24096), and written informed consent was obtained from each participant.} of the full torso in both arms-up and arms-down positions. For the in-vivo task unsupervised learning is applied with $\lambda=10^{-4}$, to both Flow U-Net and the $\zeta$ propagation variant of SVFlowNet. To evaluate the compressibility of the optimal deformations obtained using both Flow U-Net and SVFlowNet, we count the number of regions for which self-folding occurs ($\varepsilon_\textrm{reg}$)  i.e. where the Jacobian determinant $\operatorname{det}(J_{\phi})\leq0$. Such an accumulated self-folding over the image is defined as the number of voxels where the tissue collapses with itself, i.e.,
\begin{equation}\label{eqn:chapter7/field_regularity}
    \varepsilon_{reg}(\phi) = \sum_{\mathbf{x} \in \Omega} F_{\phi}(\mathbf{x})
\end{equation} 
where $F$ is the folding at a given voxel $\mathbf{x}$ defined as
\begin{equation} 
F_{\phi}(\mathbf{x}) = 
\begin{cases} 
      1 & \operatorname{det}(J_{\phi})(\mathbf{x}) \leq 0 \\
      0 & \text{elsewhere.} 
   \end{cases}
\end{equation} 
Besides field compressibility (i.e. value of Jacobian determinant) and self-folding, we also require $f_1 \circ \tilde{\phi} \approx f_0$, i.e. the registered image to approximate its pair. Therefore, we evaluate the image error ($\varepsilon_\text{img}$) using the NCC. 
 
\begin{figure}[t]
\begin{center}
\includegraphics[width=0.75\textwidth]{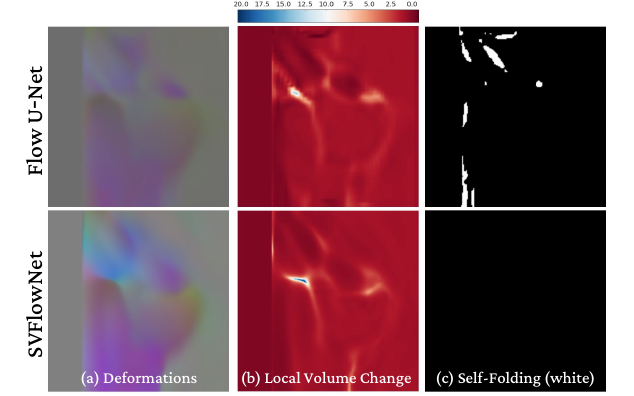}
\end{center} 
\caption{Optimal deformations (a) determined by SFlowNet and Flow U-Net for the unsupervised arms up and down breast MR image task. The Jacobian Determinant (b) and Folding Map (c) show SVFlowNet with zero self-folding compared to Flow U-Net which exhibits regions of self-folding. Data sampled from a sagittal plane. The displacement field is visualised by normalising each component and encoding the anterior, lateral and cranial directions in the red, green and blue channels respectively.}
\label{fig:invivo_results_flow}
\end{figure}

The results of the registration preserved and aligned the anatomical structures of the nipple, pectoral muscle and fibroglandular tissue, without introducing image artifacts (see Figure \ref{fig:invivo_results_flow}). On the evaluation of the optimal deformations (see Table \ref{tab:invivo_results}), Flow U-Net and SVFlowNet yield similar accurate results with an NCC error of $0.973$ and $0.968$ respectively. These are the optimal deformations in the sense of the loss function used in the optimisation performed by the Stochastic Gradient Descent performed during the training phase of the neuronal network. In this context, "optimal" means that the deformation reduces intensity mismatch after registration while avoiding compressible behaviour in the tissue. Although Flow U-Net and SVFlowNet have similar performance in terms of $\varepsilon_\text{img}$, the deformation predicted by Flow U-Net contains regions of self-folding (see Table \ref{tab:invivo_results}). This behaviour is incompatible with the deformations of breast tissue as the tissue cannot vanish or interpenetrate itself. On the other hand, SVFlowNet achieves similar performance on the image similarity and uses diffeomorphic constraints avoiding the previous non-physical behaviours yielding, as a consequence, an invertible deformation, i.e., $\operatorname{det}(J_{\phi})(\mathbf{x}) > 0, \, \forall \mathbf{x} \in \Omega$. See Appendix D for more visual results.

\begin{table}[!htp]
    \begin{center}
    \begin{tabular}{ |c|c|c|c| } 
    \hline
    Method & $\varepsilon_\textrm{reg}(\Tilde{\phi})$ & $\varepsilon_\textrm{img}(f_0, f_1 \circ \Tilde{\phi})$ \\
    \hline
    Flow U-Net & 219257 &  0.973\\ 
    SVFlowNet & 0  & 0.968  \\ 
    \hline
    \end{tabular}
    \end{center}
    \caption{Evaluation of in-vivo arms up and down MR image registration}
    \label{tab:invivo_results}
\end{table}

% \begin{figure}[!htp]
% \centering
% \includegraphics[width=\linewidth]{figures/InvivoResults.pdf}
% \caption{Transverse view of arms up (a) and down (b) breast MR images, and false colour composite predictions from SVFlowNet (c) and Flow U-Net (d).}
% \label{fig:invivo_image_pred_sag}
% \end{figure} 

% \begin{wrapfigure}{R}{5.5cm}
% \includegraphics[width=6.5cm]{figures/Invivo_results_flow.pdf} 
% \caption{}\label{fig:invivo_results_flow}
% \end{wrapfigure}  

\section{Conclusion}  
This work has presented learning strategies for breast MR image registration by introducing SVFlowNet, a novel network architecture that integrates diffeomoprhic constraints into a dual stream pyramid registration network architecture \cite{Young2022}. We have demonstrated that our method of propagating diffeomorphic deformation via the $\zeta$ operator (BCHD) outperforms the common method of summing SVFs \cite{Mok2020}, and outperforms the state-of-the-art non-diffeomorphic baseline on an in-silico breast MR image unsupervised registration task (See Appendix C). Furthermore, by construction SVFlowNet complies with underlying breast biomechanics which enforces mass conservation. For such a task, we use a measure of the ground-truth and optimal deformation discrepancy in our analysis of the performance.

On the in-vivo breast MR image arms up and down unsupervised registration task, we showed that the $\zeta$ variant of SVFlowNet achieves similar image accuracy with a state-of-the-art non-diffeomorphic baseline Flow U-Net. With SVFlowNet, we achieve an image alignment with a normalised cross-correlation ($\varepsilon_\text{img}$) of $0.968$ with zero self-folder. This is an improvement on the non-diffeomorphic baseline, for which $\varepsilon_\textrm{reg}\approx 2 \times 10^5$ have a Jacobian determinant less than or equal to zero. Thus, we have demonstrated that our approach can achieve image alignment similar to a state-of-the-art non-diffeomorphic baseline while simultaneously reproducing a physically valid estimation.

\section*{Acknowledgments}   
The authors are grateful for financial support from the New Zealand Ministry for Business, Innovation and Employment (UOAX1004), the University of Auckland Foundation (F-IBE-BIP), and the New Zealand Breast Cancer Foundation (R1704).

\bibliographystyle{splncs04}
\bibliography{references}
\newpage  

\section*{Appendices} 

\subsection*{Appendix A: Flow U-Net Architecture} 
\begin{figure}[!htp]
    \centering
    \includegraphics[width=0.78\textwidth]{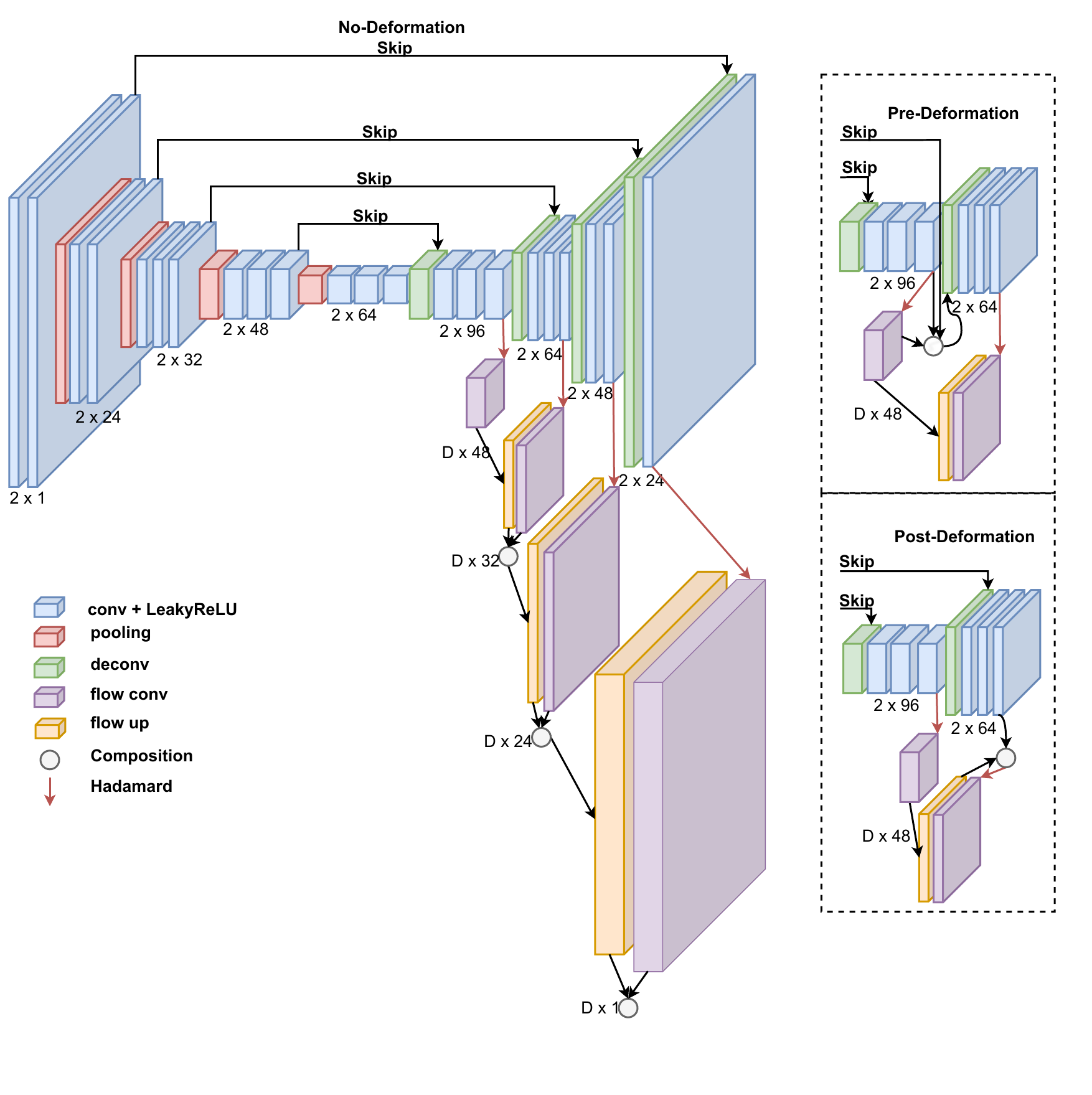}
    \caption{Flow U-Net \cite{Young2022} variants: 1. No-Deformation) Flow blocks and Hadamard transform alone, 2. Pre-Deformation) deformation applied to output feature maps of the previous $l-1$ deconvolution, and 3. Post-Deformation) deformation applied to output feature maps of the current $l$ deconvolution block.}
    \label{fig:flowunet}
\end{figure}

The Pre-Deformation variant of Flow U-Net (Fig \ref{fig:flowunet}) is used in the in-silico (Section 4.1) and in-vivo (Section 4.2) experiments for comparison with SVFlowNet. All variants (Fig \ref{fig:flowunet}) are used in the ablation study (Appendix C).

\subsection*{Appendix B: BCHD Layer}
$\zeta$ denotes the BCHD series  \cite{Mger2019NotesOT} truncated after the forth order term, 
\begin{align} 
    \nonumber
    \zeta: &(\mathbf{v}^{(l-1)}, \mathbf{v}^{(l)}) \mapsto \mathbf{v}^{(l-1)} + \mathbf{v}^{(l)} + \frac{1}{2}[\mathbf{v}^{(l-1)}, \mathbf{v}^{(l)}] ...  \\ &+ \frac{1}{12}([\mathbf{v}^{(l-1)}, [\mathbf{v}^{(l-1)}, \mathbf{v}^{(l)}]]  + [\mathbf{v}^{(l)}, [\mathbf{v}^{(l)}, \mathbf{v}^{(l-1)}]])  - \frac{1}{24}[\mathbf{v}^{(l)}, [\mathbf{v}^{(l-1)}, [\mathbf{v}^{(l-1)}, \mathbf{v}^{(l)}]]] .
\end{align}  

\subsection*{Appendix C: In-Silico Ablation Study}

\begin{figure}
    \centering
    \includegraphics[width=\textwidth]{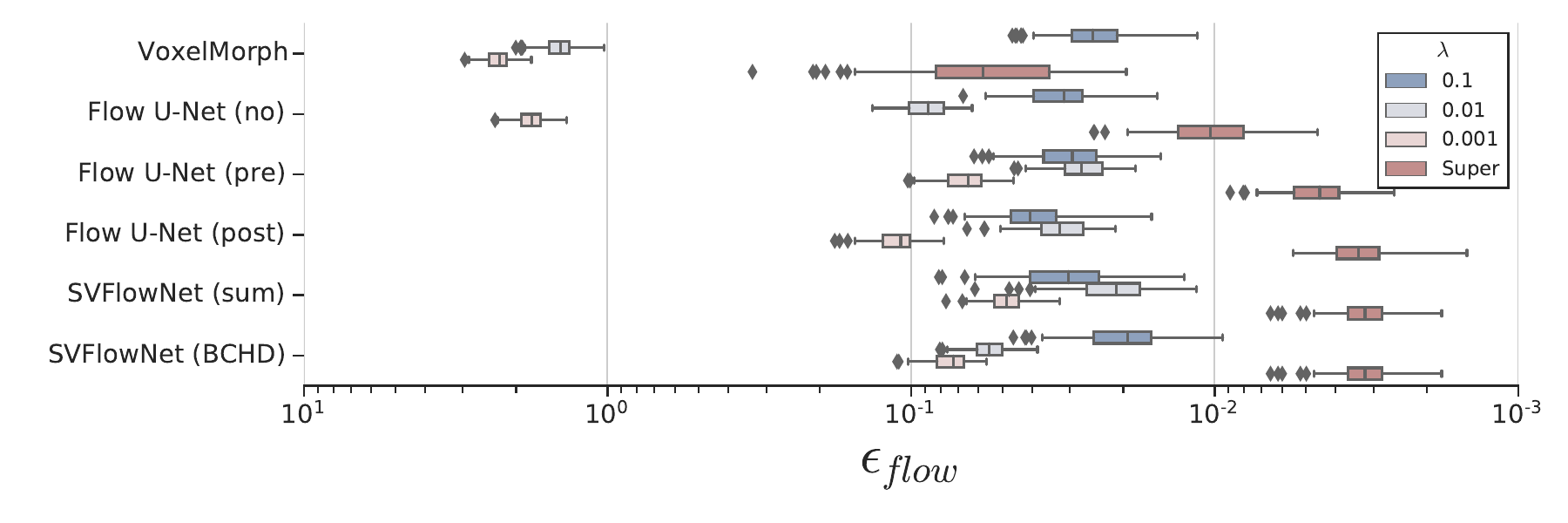}
    \caption{Performance of all methods on the in-silico data assessed using the deformation discrepancy $\epsilon_\textrm{flow}$.}
    \label{fig:ablation_insilico}
\end{figure}

We performed the same in-silico experiment in Section 4.1 using VoxelMorph \cite{balakrishnan2018}, the Flow U-Net variants \cite{Young2022} (Appendix A, Fig \ref{fig:flowunet}) and the SVFlowNet variants. The results are compared in Fig \ref{fig:ablation_insilico}.

\subsection*{Appendix D: In-Vivo Results Visualization}   
\begin{SCfigure}
    \centering
    \includegraphics[width=0.7\textwidth]{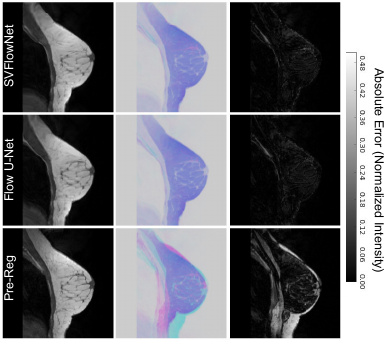}
    \caption{In-vivo pre- and post-registered images: predicted arms up images (left), false color composites (center), and voxel-wise absolute errors (right).}
    \label{fig:vis_invivo}
\end{SCfigure}

A visualisation of the image reconstruction results from the in-vivo experiment (Section 4.2) is shown in Fig \ref{fig:vis_invivo}.
% \begin{figure}[htp!]%
% \centering
% \subfigure[Predicted Arms Up]{%
% \label{fig:first}%
% \includegraphics[width=\textwidth]{figures/vis_img.pdf}}%
% \qquad
% \subfigure[Composite]{%
% \label{fig:second}%
% \includegraphics[width=\textwidth]{figures/vis_comp.pdf}}%  
% \qquad
% \subfigure[Absolute Error (Voxel-Wise)]{%
% \label{fig:third}%
% \includegraphics[width=\textwidth]{figures/vis_diff.pdf}}%
% \caption{VoxelMorph \cite{balakrishnan2018}, Flow U-Net variants \cite{Young2022} (illustrated in Appendix A) and SVFlowNet variants performance on the in-silico data assessed using the deformation discrepancy $\epsilon_\textrm{flow}$ and image similarity $\epsilon_\textrm{image} = 1 - NCC(f_0, f_1 \circ \phi)$.} 
% \label{fig:ablation_insilico}
% \end{figure}    

\end{document}